\baselineskip 18pt
\magnification=\magstep1
\hsize=15.0truecm
\hoffset=1.0truecm
\vsize=22.0truecm

\def\spc{\hskip 3 pt}

\hfill {\vbox {\hbox {BARI-TH 208/95} \hbox {\it November 1995}}}
\vskip 1truecm
{\centerline {\bf GENERALIZED GAUSSIAN EFFECTIVE POTENTIAL: }}
\vskip .2 truecm
{\centerline {\bf SECOND ORDER THERMAL CORRECTIONS}}
\vskip 0.7truecm
{\centerline {Paolo Cea \spc\spc\spc and \spc \spc \spc Luigi Tedesco}}
\vskip 0.5truecm
{\it Dipartimento di Fisica dell'Universit\'a di Bari, I-70126 Bari, Italy
\par
{\centerline {INFN, Sezione di Bari, I-70126 Bari, Italy}}}
\vskip 2.3truecm

{\centerline {\bf ABSTRACT}}

\vskip 1truecm
\par\noindent
We discuss the finite temperature generalized
Gaussian effective potential. We put out a 
very simple relation between the thermal corrections to the generalized
Gaussian effective potential  
and those of the  effective potential. We evaluate explicitly 
the second order thermal corrections in the case of the
selfinteracting scalar field in one spatial dimension.

\vfill\eject
In a previous paper [1] (henceforth referred to as I) we investigated the 
thermal corrections to the generalized Gaussian effective potential [2].
The starting point was the set-up of a 
perturbation theory with a variational basis
which allowed to evaluate in a systematic way the corrections to the 
variational Gaussian approximation. In Ref. [2] we showed that the Hamiltonian 
$H$ of a selfinteracting scalar field can be naturally decomposed into a free 
term $H_0$ and an interacting term $H_I$. It turns out that $H_0$ is the hamiltonian 
of a free scalar field with mass $\mu(\phi_0)$ satisfying the so called gap 
equation (in $\nu$ spatial dimensions)

$$
\mu^2=m^2 +{{\lambda}\over 2} \phi_0^2 +
{{\lambda}\over 4}\int {{d^{\nu} k}\over {(2 \pi)^{\nu}}} {1\over 
{\sqrt{{\vec{k}}^2+\mu^2}}}.\eqno(1)
$$

\vskip 0.5truecm
\par\noindent
The interaction $H_I$ is given by the off-diagonal elements of the full
Hamiltonian $H$ with respect to the variational basis. Moreover, we 
showed that

$$
H_I=\int d^{\nu}x \left[ \left(\mu^2 \phi_0 - {\lambda\over 3} \phi^3_0 \right)
:\eta(\vec {x}): + {\lambda\over {3!}} \phi_0 : \eta^3(\vec {x}): + 
{\lambda\over {4!}} :\eta^4(\vec {x}): \right],\eqno(2)
$$

\vskip 0.5truecm
\par\noindent
where $\eta(\vec{x})$ is the fluctuating field, and the colon denotes 
normal 
ordering with respect to the variational ground state.
\par\noindent
In I we evaluated the thermal corrections to the generalized Gaussian 
effective potential $V_G^T$
by means of the standard thermodinamic perturbation
theory [3]. In this way we obtained

$$
V_G^T(\phi_0)=V_{GEP}(\phi_0)+{1\over {\beta}}\int {{d^{\nu} k}\over 
{(2 \pi)^{\nu}}} ln \left(1-e^{-\beta g(\vec{k})}\right) + \Delta V_G^T(\phi_0)
\eqno(3)
$$

\vskip 0.5truecm
\par\noindent
where $g(\vec{k})=\sqrt{{\vec{k}}^2 + \mu^2(\phi_0)}$ and ($V$ is the spatial 
volume)

$$\Delta V_G^T(\phi_0)=-{1\over {\beta V}}\sum_{m=2}^{\infty}
{{(-1)^m}\over {m!}}
\int_0^{\beta} d \tau_1...\tau_m<T_{\tau}(H_I(\tau_1)...H_I(\tau_m)>^{\beta}_c.
\eqno(4)
$$

\vskip 0.5truecm
\par\noindent
In Equation (4) $H_I(\tau)$ is the interaction Hamiltonian in the Matsubara 
interaction picture [4], and the thermal average is done with respect the free  
Hamiltonian $H_0$.
\par
The aim of the present paper is to discuss some consequences of the general 
formulae (3) and (4). Moreover we evaluate the second order thermal corrections 
in the case of selfinteracting scalar field in one spatial dimension.
\par\noindent
In the lowest order approximation, the thermal generalized Gaussian effective 
potential is given by Eq. (3) with $\Delta V_G^T(\phi_0)=0$.
It is interesting to compare our result with the finite temperature Gaussian 
effective potential discussed by Hajj and Stevenson [5] (see also Ref. [6]).
\par\noindent
In order to evaluate the thermal corrections to the Gaussian effective 
potential within a non-perturbative approach, the authors of Ref. [5] decomposed 
the Hamiltonian into two terms:

$$
H=H_0+H_I, \eqno(5)
$$

\vskip 0.5truecm
\par\noindent
where $H_0$ is the Hamiltonian of free scalar particle with variational mass 
$M$. Whereupon one evaluates the free energy by using the standard 
thermodinamic perturbation theory. Hajj and Stevenson, after evaluating
the thermodinamic potentials up to the first order  in the perturbation $H_I$,
fix the variational mass $M$ by minimizing the free energy density. 
As a consequence the mass $M$ satisfies a gap equation which includes the 
thermal corrections. This is the main difference  between our approach and
the one of Ref. [5]. Indeed in our approach the variational basis is fixed once 
and for all at $T=0$ by Eq. (1).
We stress that  in Ref. [5] the thermal  variational mass and the interaction 
Hamiltonian depend
on the approximation adopted in evaluating the free energy density.
On the other hand, in our approach the interaction Hamiltonian is determined
by the variational basis at T=0.
\par
It is worthwile to compare our lowest order thermal corrections with 
the 1-loop
finite temperature effective potential  [7,8]. In the one loop approximation 
the finite temperature effective potential is well known:

$$
V^1_{\beta}(\phi_0)={1\over {2 \beta}} \sum_n \int {{d^{\nu}k}\over 
{(2 \pi)^{\nu}}} ln (E^2+ \omega_n^2) \eqno(6)
$$

\vskip 0.5truecm
\par\noindent
where $\omega_n={{2 \pi n}\over {\beta}}$, $E^2= {\vec{k}}^2 + M^2(\phi_0)$,
and 
$M^2(\phi_0)=m^2+{{\lambda}\over 2} \phi_0^2$. The sum over $n$ can be readily
evaluated [7]:

$$
V^{(1)}_{\beta}(\phi)=\int {{d^{\nu} k}\over {(2 \pi)^{\nu}}} {E\over 2}+
{1\over {\beta}} \int {{d^{\nu} k}\over {(2 \pi)^{\nu}}} ln (1-e^{-\beta E}).
\eqno(7)
$$

\vskip 0.5truecm
\par\noindent
The first term in Eq. (7) is the zero temperature 1-loop effective potential.
Thus the 1-loop thermal correction is given by the second term in (7).
Now, comparing Eq. (7) with Eq. (3) we see that 1-loop thermal correction to 
the effective potential coincides with the lowest order thermal correction to 
the generalized Gaussian effective potential if

$$
M^2(\phi_0)=m^2+{{\lambda}\over 2} \phi_0^2 \, \,
\longrightarrow \,\, \mu^2(\phi_0).
\eqno(8)
$$

\vskip 0.5truecm
\par\noindent
In other words, if in the thermal correction to the effective potential we 
replace the tree level mass of the shifted theory with the mass $\mu(\phi_0)$
obtained by summing  the superdaisy graphs at $T=0$ in the propagator, then we 
obtain again a free energy density.  Up to now this remarkable result in 
thermal scalar 
field theories holds  for the lowest order correction only. 
We show, now, that it extends also to the higher order thermal corrections.
To this end 
we observe that the higher order corrections are given by Eq. (1.9) of Ref. 
[7]. On the other hand, in our approach the thermal corrections can be 
evaluated by means of Eq. (4). Observing that $L_I=-H_I$ and that the Gaussian 
functional integration with periodic boundary conditions in Ref. 
[7] corresponds
to the thermal Wick theorem, we obtain the desired result. However, because our 
interaction Hamiltonian is normal ordered, to complete the proof we must show
that the thermal corrections are not affected by the normal ordering of the 
interaction Hamiltonian. To see this, we note that 
the normal ordering is ineffective when we consider
a thermal contraction of of two scalar fields belonging to different vertices.
Therefore, the normal ordering comes into play when we contract two fields 
which belong to the same vertex. In this case we get the following thermal 
average

$$
{\tilde{G}}_{\beta}(0)=<T_{\tau}:\eta(\vec{x},\tau) \eta(\vec{x},\tau):>
^{\beta}, \eqno(9)
$$

\vskip 0.5truecm
\par\noindent
instead of $G_{\beta}(0)$, where $G_{\beta}(\vec{x},\tau)$ is the thermal 
propagator:

$$G_{\beta}(\vec{x},\tau)=
<T_{\tau} \eta(\vec{x},\tau) \eta(0)>^{\beta}=
{1\over {\beta}} \sum_{n=-\infty}^{+\infty}
 \int {{d^{\nu} k}\over {(2 \pi)^{\nu}}}
 { {e^{i [{\vec{k}} \cdot \vec{x}-\omega_n \tau]
}}\over {\omega_n^2+g^2(\vec{k})}}.\eqno(10)
$$

\vskip 0.5truecm
\par\noindent
Taking into account the canonical commutation relations between the creation 
and annihilation operators, it is straighforward to show that:

$$
{\tilde{G}}_{\beta}(0)= G_{\beta}(0)- \int {{d^{\nu} k}\over {(2 \pi)^{\nu}}}
{1\over {2 g(\vec{k})}}.\eqno(11)
$$

\vskip 0.5truecm
\par\noindent
Now we observe that 

$$
\int {{d^{\nu} k}\over {(2 \pi)^{\nu}}} {1\over {2 g(\vec{k})}} =
\lim_{\beta \to \infty} G_{\beta}(0).\eqno(12)
$$

\vskip 0.5truecm
\par\noindent
Indeed, from Eq. (10) it follows:

$$
G_{\beta}(0) = {1\over {\beta}} \sum_{n=-\infty}^{+\infty} \int
{{d^{\nu} k}\over {(2 \pi)^{\nu}}} {1\over {\omega^2_n + g^2(\vec{k})}}.
\eqno(13)
$$

\vskip 0.5truecm
\par\noindent
By using the identity

$$
{\rm cotgh}(x)={1\over {\pi x}}+{{2x}\over {\pi}}\sum_{n=1}^{\infty}
{1\over {x^2+n^2}}, \eqno(14)
$$

\vskip 0.5truecm
\par\noindent
we  rewrite Eq. (13) as:

$$
G_{\beta}(0)=\int {{d^{\nu} p}\over {{(2 \pi)}^{\nu}}} {1\over {2 g(\vec{p})}}
{\rm cotgh} \left[{{\beta g(\vec{p})}\over 2}\right]. \eqno(15)
$$

\vskip 0.5truecm
\par\noindent
Finally, performing the limit $\beta \rightarrow \infty$ in Eq. (15) we obtain 
Eq. (12). 
Thus, we have shown that the normal ordering of the interaction Hamiltonian 
does not modify the thermal corrections.
\par\noindent
Note that from Eqs. (11) and (15) it follows that

$$
{\tilde{G_{\beta}}}(0)=\int { {d^{\nu} k}\over {(2 \pi)^{\nu}}} 
{1\over {2 g(\vec{k})}} \left[
{\rm cotgh} \left({{\beta}\over 2} g(\vec{k})\right)-1\right].
\eqno(16)
$$

\vskip 0.5truecm
\par\noindent
Equation (16) shows that ${\tilde{{G}_{\beta}}}(0)$ is finite 
for any value of $\nu$. 
\par
Let us, now, evaluate the second order thermal 
corrections in the case of one spatial dimension, $\nu=1$. From Eq. (4)
we have

$$
\Delta V_G^T (\phi_0)=-{1\over {2! \, \beta V}} \int_0^{\beta} d \tau_1 
d \tau_2 < T_{\tau} H_I(\tau_1) H_I(\tau_2)>^{\beta}_c. \eqno(17)
$$

\vskip 0.5truecm
\par\noindent
In Figure 1 we display the second oder thermal corrections obtained with the 
aid of the thermal Wick theorem. The solid lines correspond to the thermal 
propagator Eq. (10), the vertices can be extracted from the interaction 
Hamiltonian Eq. (2). Let us analyze the graphs in Fig. 1.
It is easy to see that graph (a) is temperature-independent. 
So it does not contribute
to $\Delta V_G^T$ due to the stability condition $<\Omega\,|\, \eta\,|\,\Omega>
=0$. As concern the graph (b), we have

$$
\eqalign{
(b)=-{{\lambda \phi_0}\over {4 \beta V}}  \left(\mu^2 \phi_0-{{\lambda}\over 3}
\phi_0^3\right) \int_{-\infty}^{+ \infty} dx \, dy \int_0^{\beta} d \tau_1 \,
d \tau_2 & <T_{\tau} \eta(x, \tau_1) \eta(y, \tau_2)>^{\beta} \cr
 & <T_{\tau} :\eta(y, \tau_2) \eta(y, \tau_2):>^{\beta}.\cr}
$$

\vskip 0.5truecm
\par\noindent
According to our previous discussion we obtain

$$
(b) = -{1\over 4}\lambda  \phi_0^2 \left(1-{{\lambda}\over {3 \mu^2}}\phi_0^2
\right) 
{\tilde{G}}_{\beta}(0).\eqno(18)
$$

\vskip 0.5truecm
\par\noindent
In a similar way we find:

$$
(c)=- {{{\lambda}^2 \phi_0^2}\over 8} \,\, {1\over {\mu^2}} \,\,
{\tilde{G}}_{\beta}^2(0).\eqno(19)
$$

\vskip 0.5truecm
\par\noindent
For the graphs (d) we have

$$
(d)=-{{{\lambda}^2 \phi_0^2}\over {2 \cdot 3!}} \int_{-{{\beta}\over 2}}
^{+{{\beta}\over 2}} d \tau \int_{-\infty}^{+\infty} 
dx  \,\,  G^3_{\beta}(x,\tau).
$$

\vskip 0.5truecm
\par\noindent
Using Eq. (10) and the result

$$
{1\over {\beta}}\sum_n{{e^{i \omega_n \tau}}\over {\omega_n^2+g^2(k)}}= 
{{e^{- g(k) |\tau|}}\over {2 g(k)}}+{1\over {2 g(k)}}
 {{(e^{g(k) \tau} +
e^{- g(k) \tau})}\over {e^{\beta g(k)}-1}},\eqno(20)
$$

\vskip0.5truecm
\par\noindent
we get

$$
(d)= - {{{\lambda}^2 \phi_0^2 }\over {48 (2 \pi)^2}}
 \int_0^{\beta\over 2} d \tau 
\int_{-\infty}^{+\infty}  {{d k_1 \, dk_2}
\over {g(k_1)g(k_2)g(k_3)}}
\prod_{i=1}^3 \left[e^{-g(k_i) \tau} + { {e^{g(k_i)\tau} + e^{-g(k_i) \tau}}
\over {e^{\beta g(k_i)} -1}}\right]\eqno(21)
$$

\vskip 0.5truecm
\par\noindent
where $\sum_{i=1}^3k_i=0$.

\par\noindent
Finally, using Eq. (20) we get:

$$
(e)=- {{\lambda^2 }\over {32 (2 \pi)}} {\tilde{G}}^2_{\beta}(0)
\int_0^{{\beta}\over 2} d \tau \,\, 
\int_{-\infty}^{+\infty} {{d k}\over {g^2(k)}} 
\left[e^{-g(k) \tau} + { {e^{g(k)\tau} + e^{-g(k) \tau}}
\over {e^{\beta g(k)} -1}}\right]^2\eqno(22)
$$

\vskip 0.5truecm
\par\noindent
and

$$\eqalignno{ 
(f)=- {{\lambda^2 }\over {16 \cdot 4! (2 \pi)^3}}\int_0^{{{\beta}\over 2}}
d \tau & \int_{-\infty}^{+\infty} 
{{d k_1 d k_2 d k_3 }\over 
{g(k_1) g(k_2) g (k_3) g(k_4)}}\cdot \cr
 & \prod_{i=1}^4  
\left[e^{-g(k_i) \tau} + { {e^{g(k_i)\tau} + e^{-g(k_i) \tau}}
\over {e^{\beta g(k_i)} -1}}\right]&(23)\cr}
$$

\vskip 0.5truecm
\par\noindent
with $\sum_{i=1}^{4} k_i=0$.
\par\noindent
A few comments are in order. 
In Equations (21), (22) and (23) the $\tau$-integration can be performed
explicitly, while the remaining integrations over the momenta $k_i$ must be 
handled numerically.
In the limit $\beta \rightarrow \infty \,
(T \rightarrow 0)$ the anomalous "graphs" (b), (c) and (e) go to zero
exponentially due to the factor ${\tilde {G}}_{\beta}(0).$
On the other hand, the graphs (d) and (f) reduce to the zero 
temperature second order corrections to the Gaussian effective potential [9].
Indeed, in that limit in Eqs. (21-23)
only the factor $e^{-g(k_i)\tau}$ survives. 
Performing the elementary $\tau$-integration we obtain the zero 
temperature contributions.
As a consequence the zero temperature limit of $V_G^T(\phi_0)$ reduces to
$V_G(\phi_0)$.
\par\noindent
In the high temperature limit $\beta \rightarrow 0$ we find that the graphs (e)
and (f) dominate. Therefore, in the intermediate temperature region 
$\beta \sim 1$ we expect that the main contribute to $V_G^T(\phi_0)$
comes from the graphs (d), (e) and (f). Indeed this is the case as shown
in Fig. 2.
\par\noindent
In Figure 3 we display the finite temperature generalized Gaussian
effective potential for three different values of $T$ and 
$\hat{\lambda}>{\hat{\lambda}}_c \simeq 1.15$ [9].
Figure 3 shows that the symmetry broken at $T=0$ is restored by increasing the 
temperature through a continuous phase transition.
\par
In conclusion, in this paper we have discussed the thermal corrections to the 
generalized Gaussian effective potential. 
Remarkably, we found that the thermal corrections can be obtained from those of 
the effective potential
with the substitution Eq. (8). 
Moreover, we have evaluated the second order corrections in the case of 
selfinteracting scalar fields in one spatial dimension. We plan to extend
our work to the case of higher spatial dimensions in a future investigation.

\vfill\eject
{\bf FIGURE CAPTIONS}
\vskip 1truecm
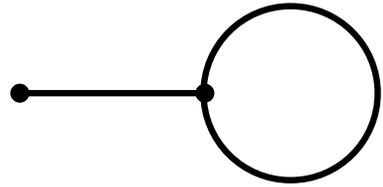
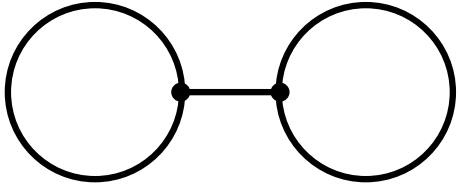
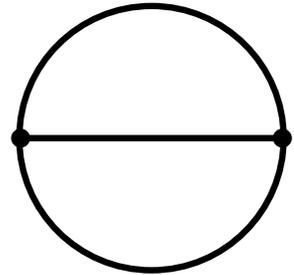
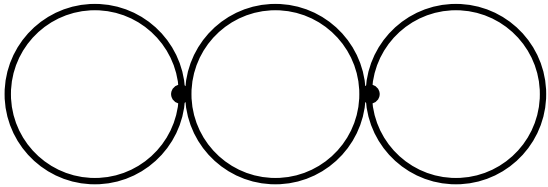
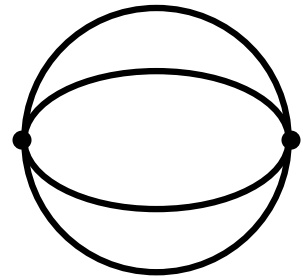
\item {\bf {Figure 1}} Second order thermal corrections to the generalized 
Gaussian effective potential.
\vskip 0.5truecm
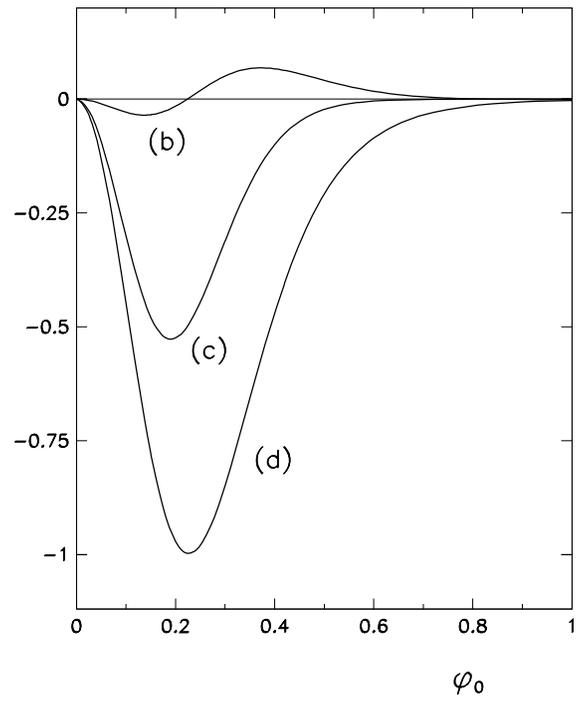
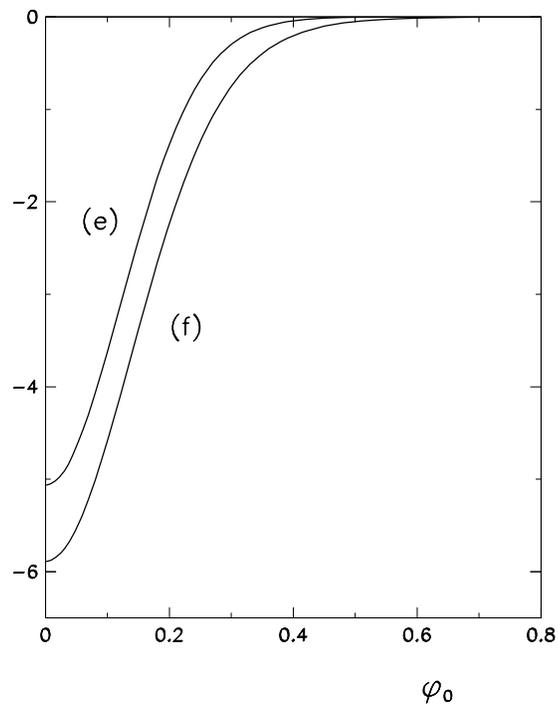
\item {\bf {Figure 2}} Contributions to $V_G^T(\phi_0)$ due to the graphs in 
Fig. 1 for $\hat{\lambda}=4$ and $\hat{T}=1$ (notations as in I).
\vskip 0.5truecm
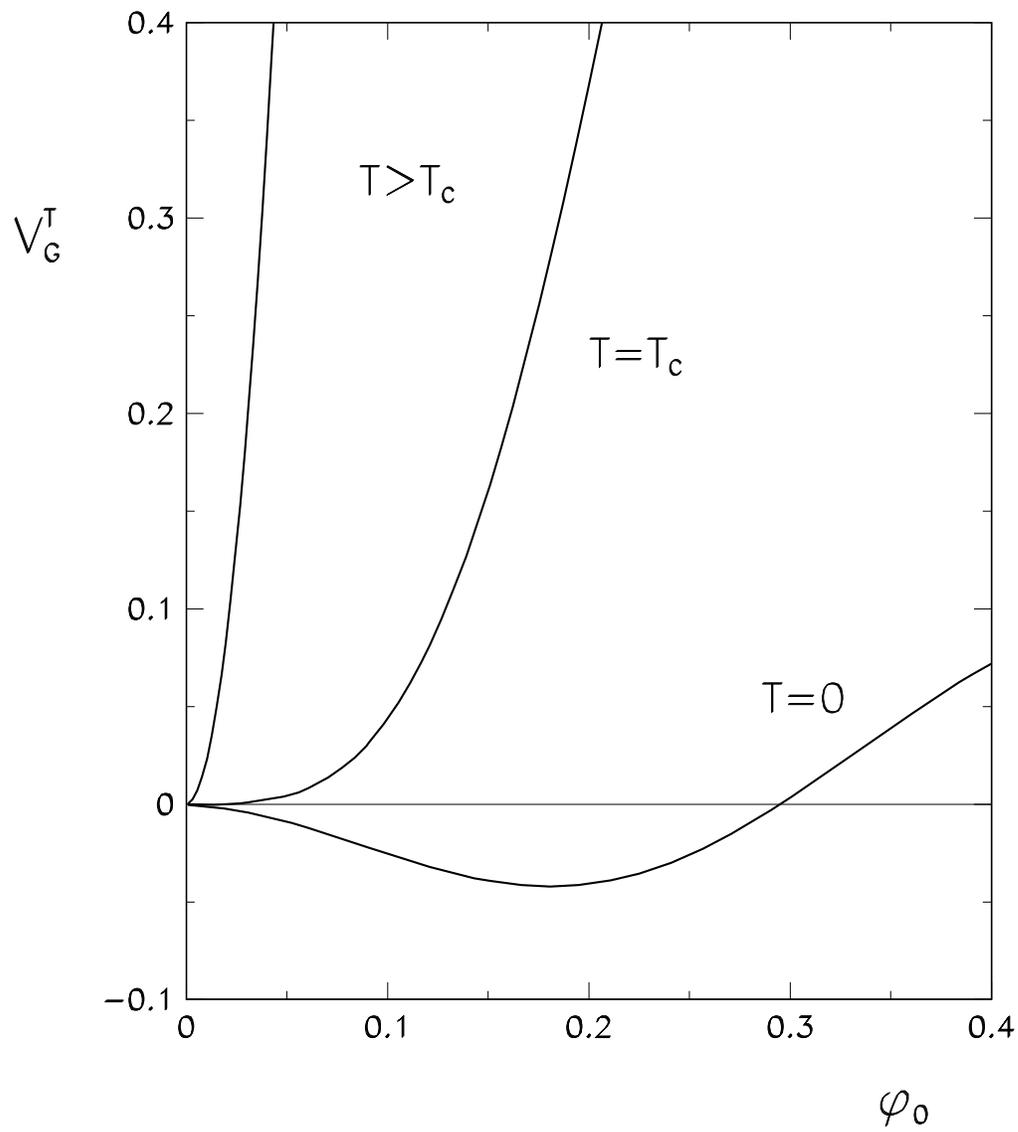
\item {\bf {Figure 3}} $V_G^T(\phi_0)$ in units of $\mu_0=\mu(\phi_0=0)$ for
$\hat{\lambda}=4$ and $\hat{T}=0$, $\hat{T}={\hat{T}}_c \simeq 0.764$, and
$\hat{T}>{\hat{T}}_c$.

\vfill\eject
{\bf REFERENCES}
\vskip 1truecm
\item {[1]} {P. Cea and L. Tedesco, 
{\it Finite Temperature Generalized Gaussian 
Effective Potential}, BARI-TH/198/95.
\item {[2]}  P. Cea, {\it Phys. Lett.} {\bf {B 236}} (1990) 191;
\par
P. Cea and L. Tedesco, {\it Phys. Lett.} {\bf B 335} (1994) 423.
\item {[3]} See, for istance, A.A. Abrikosov, L.P. Gorkov, and I.E. 
Dzyaloshinski, {\it Methods of Quantum Field Theory in Statistical Physics}
(Dover, New York, 1975);
\par
A.L. Fetter and J.D.Walecka, {\it Quantum Theory of Many Particle System}
(Mc 
\par
Graw-Hill, New York, 1965).
\item {[4]} T. Matsubara, {\it Prog. Theor. Phys.} {\bf 177} (1955) 351.
\item {[5]} G.A. Hajj and P.M. Stevenson, {\it Phys. Rev.} {\bf D 37}
(1988) 413.
\item {[6]} I. Roditi, {\it Phys. Lett.} {\bf B 177} (1986) 85;
\par
A. Okopinska, {\it Phys. Rev.} {\bf D 36} (1987) 2415.
\item {[7]} L. Dolan and R. Jackiw, {\it Phys. Rev.} {\bf D 9} (1974) 3320.
\item {[8]} S. Weinberg, {\it Phys. Rev.} {\bf D 9} (1974) 3357.
\item {[9]} L. Tedesco, {\it Ph. D. Thesis}, University of Bari, 1995 
(unpublished).

\bye